\documentclass[a4paper]{article}

\usepackage{INTERSPEECH2022}
\usepackage{CJKutf8}

\title{\textbf{A Polyphone BERT for Polyphone Disambiguation in Mandarin Chinese}}
\name{\textit{Song Zhang, Ken Zheng, Xiaoxu Zhu, Baoxiang Li}}
\address{SenseTime Research}
\email{\{zhangsong1, zhengken, zhuxiaoxu, libaoxiang\}@sensetime.com}
\usepackage{cite}
\usepackage{enumerate}

\begin{document}

\maketitle
\begin{abstract}
  Grapheme-to-phoneme (G2P) conversion is an indispensable part of the Chinese Mandarin text-to-speech (TTS) system, and the core of G2P conversion is to solve the problem of polyphone disambiguation, which is to pick up the correct pronunciation for several candidates for a Chinese polyphonic character. In this paper, we propose a Chinese polyphone BERT model to predict the pronunciations of Chinese polyphonic characters. Firstly, we create 741 new Chinese monophonic characters from 354 source Chinese polyphonic characters by pronunciation. Then we get a Chinese polyphone BERT by extending a pre-trained Chinese BERT with 741 new Chinese monophonic characters and adding a corresponding embedding layer for new tokens, which is initialized by the embeddings of source Chinese polyphonic characters. In this way, we can turn the polyphone disambiguation task into a pre-training task of the Chinese polyphone BERT. Experimental results demonstrate the effectiveness of the proposed model, and the polyphone BERT model obtain 2\% (from 92.1\% to 94.1\%) improvement of average accuracy compared with the BERT-based classifier model, which is the prior state-of-the-art in polyphone disambiguation.
\end{abstract}

\section{Introduction}

The TTS system plays an important role in the human-computer voice interaction framework, which aims to translate text into speech. In a mandarin TTS system, G2P is a task that changes Chinese natural language texts into Pinyins. For most Chinese characters, the pronunciation is fixed. However, there are some special cases, the pronunciation is manifold, i.e. polyphonic characters. They have different pronunciations in different contexts, and if an inaccurate pronunciation was selected, it would have a huge impact on the semantics and user experience. Therefore, identifying a correct pronunciation for a polyphonic character according to its context, called polyphone disambiguation, is an important task for the mandarin TTS system.

A variety of approaches have been proposed to address the polyphone disambiguation problem. They can be categorized into knowledge-based and learning-based approaches.

\subsection{Knowledge-based approaches}

Early studies\cite{gou1processing,zhang2002efficient,liu2011polyphone} of polyphone disambiguation mainly relied on dictionaries and rules. A well-designed dictionary and human rules are essential in a knowledge-based system. During runtime, segment input sentences into words, and then found the corresponding pronunciation in the dictionary. But the dictionary cannot cover all the polyphonic cases, some rules are crafted by linguistic experts to handle the corner cases.

\subsection{Learning-based approaches}

In recent years, the success of Neural Networks in various fields has enabled the application of Recurrent Neural Network (RNN) to solve the polyphone disambiguation task \cite{rao2015grapheme,yao2014spoken,huang2015bidirectional,mikolov2012context}. \cite{shan2016bi} treated polyphone disambiguation as a classification task, using a bidirectional long short-term memory (BLSTM) neural network and additional information such as part-of-speech (POS) tags to predict the pronunciation of the input polyphonic characters, which yielded good results. \cite{cai2019polyphone} used multi-level embedding \cite{song2018directional} features as input and yielded improvement on the task. \cite{zhang2020mask} used weighted softmax and modified focal loss to solve the problem of unbalanced training samples.

Self-training methods such as BERT \cite{devlin2018bert} and RoBERTa \cite{liu2019roberta} take advantage of unlabeled text data, which have brought significant performance in sentence comprehension. With the powerful character representations, pre-trained models can be coupled with simple DNN networks to do various downstream tasks. \cite{yang2019pre} was the first to use BERT in the polyphone disambiguation task and achieved better results than ever before.

\begin{figure}[b]
	\centering
	\includegraphics[width=\linewidth]{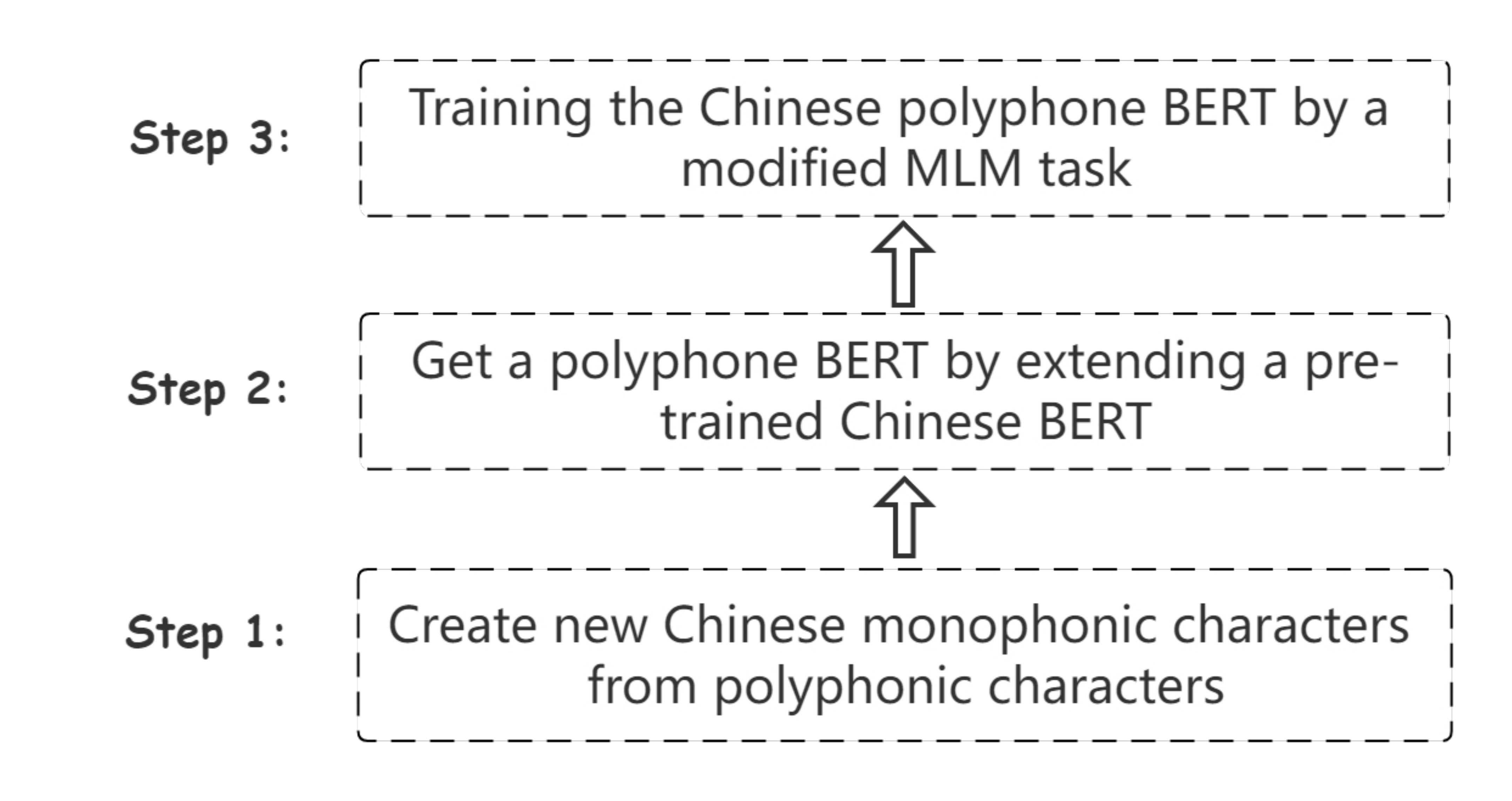}
	\caption{\textit{Steps of our proposed method}}
\end{figure}

With the success of the prompt learning in natural language processing tasks\cite{liu2021pre}, it has been proved effective to turn the downstream task into a pre-training task of the pre-train LM. Inspired by this idea, we propose a Chinese polyphone BERT, and turn the polyphone disambiguation task into a pre-training process of the polyphone BERT. The contribution of our work can be summarized into the following aspects:

\begin{itemize}
\item We propose a Chinese polyphone BERT and turn the polyphone disambiguation task into a pre-training process of the polyphone BERT.
\item Our approache model achieves state-of-the-art in Mandarin Chinese polyphone disambiguation. The proposed model can improve polyphone disambiguation accuracy on all four test sets.
\item We analyze the influence of freezing the parameters of BERT on BERT-based methods of polyphone disambiguation, which was not presented in previous works of polyphone disambiguation.
\end{itemize}

\section{Proposed method}

As shown in Figure 1, there are three steps in our method. We create new Chinese characters from polyphonic characters at first, then we get a Chinese polyphone BERT by adding the new characters into a pre-trained Chinese BERT vocabulary and initializing the weights of new tokens. Finally, we train the Chinese polyphone BERT with masked language model (MLM) layers.

\subsection{New Chinese characters}

The Chinese characters consist of monophonic characters and polyphonic characters. A Chinese polyphonic character has at least two different pronunciations, and these pronunciations are indicated by Pinyins. Different pronunciations represent different semantics for a polyphonic character. In another word, a polyphonic character can be seen as some monophonic characters using the same symbol, so we try to create new Chinese monophonic characters from Chinese polyphonic characters.

\begin{figure}[h]
	\centering
	\includegraphics[width=\linewidth]{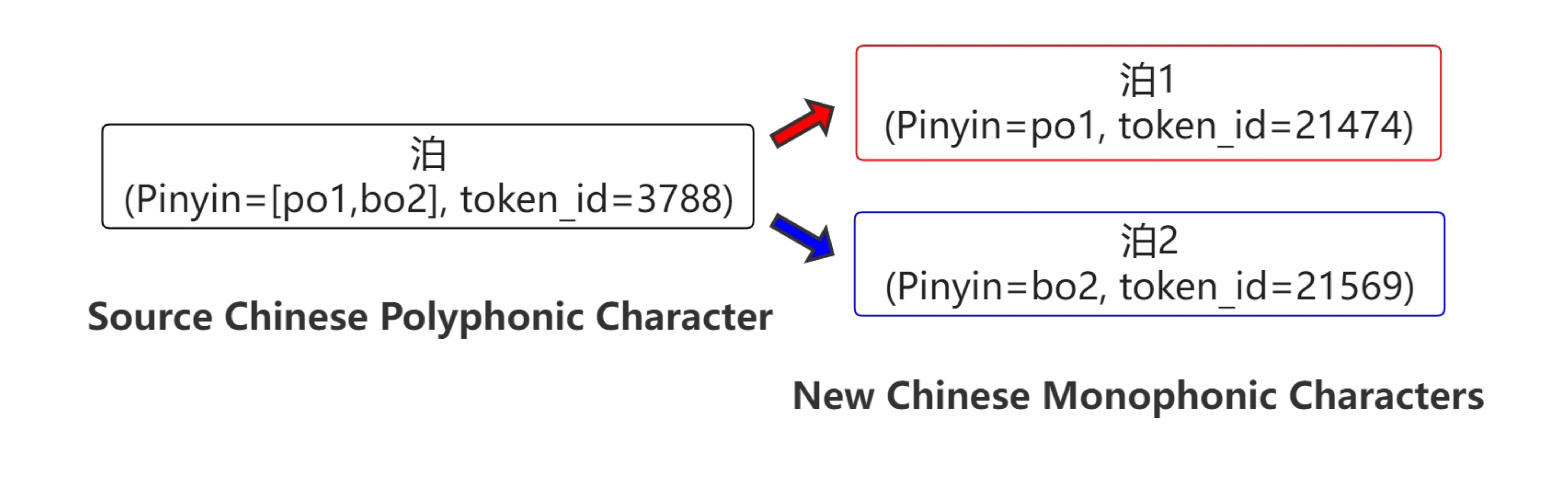}
	\caption{\textit{An example of creating new Chinese monophonic characters.}}
\end{figure}

Figure 2 is an example of creating new Chinese monophonic characters. The Chinese polyphonic character \begin{CJK}{UTF8}{gbsn}“泊”\end{CJK} has two Pinyins, \begin{CJK}{UTF8}{gbsn}“po1”\end{CJK} and \begin{CJK}{UTF8}{gbsn}“bo2”\end{CJK} , and its token id in the pre-trained Chinese BERT is 3,788. We create two new monophonic characters \begin{CJK}{UTF8}{gbsn}“泊1”\end{CJK} and \begin{CJK}{UTF8}{gbsn}“泊2”\end{CJK}. The Pinyin of \begin{CJK}{UTF8}{gbsn}“泊1”\end{CJK} is \begin{CJK}{UTF8}{gbsn}“po1”\end{CJK}, and the token id of the new character \begin{CJK}{UTF8}{gbsn}“泊1”\end{CJK} is 21,474 in the extended part of the pre-trained Chinese BERT vocabulary. The Pinyin of \begin{CJK}{UTF8}{gbsn}“泊2”\end{CJK} is \begin{CJK}{UTF8}{gbsn}“bo2”\end{CJK}, and token id of the new character \begin{CJK}{UTF8}{gbsn}“泊2”\end{CJK} is 21,569 in the extended part of the pre-trained Chinese BERT vocabulary. We can initialize the new tokens 21,474 and 21,569 by the source token 3,788 or the UNK token of the BERT. In this way, we can turn a sentence with polyphonic characters into a new sentence with no polyphonic characters if we know the Pinyins of the appeared polyphonic characters, the case of \begin{CJK}{UTF8}{gbsn} “小船漂泊在湖泊里”\end{CJK}(a boat drifted in the lake) is shown in Figure 3. The source Chinese polyphonic character in the original sentence will be written as SCPC for short, and the new Chinese monophonic Characters in the new sentence will be written as NCMC for short.

\begin{figure}[b]
	\centering
	\includegraphics[width=\linewidth]{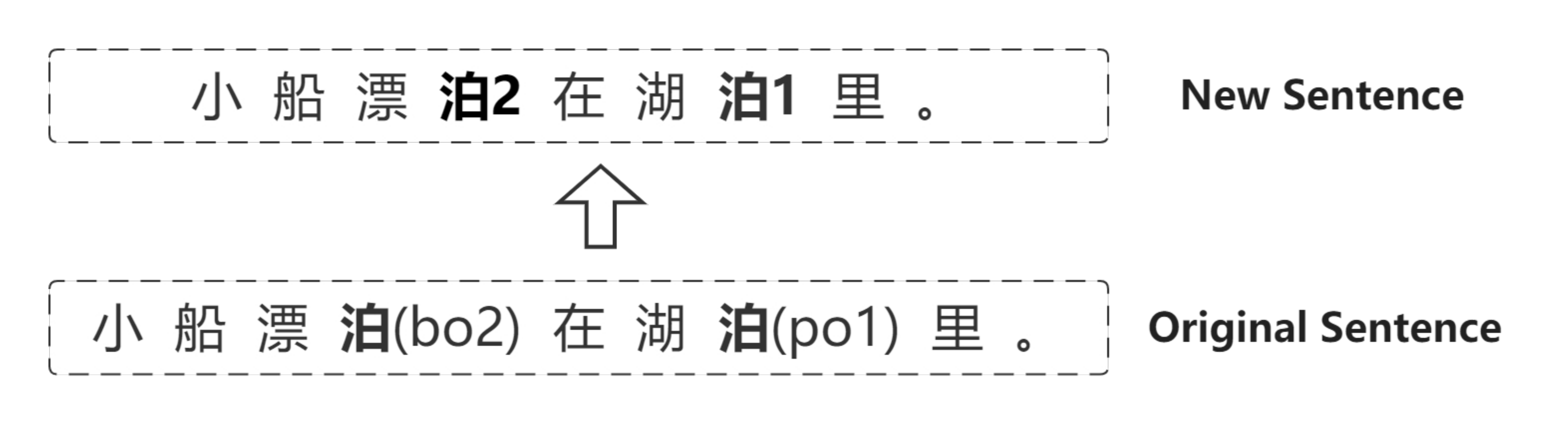}
	\caption{\textit{Turning an original sentence into a new sentence with no polyphonic characters.}}
\end{figure}

\subsection{Chinese polyphone BERT}

BERT is a deep learning Transformer model that revolutionized the way we do natural language processing. The Chinese BERT model is pre-trained on a large amount of Mandarin unlabeled data with two tasks, predicting the masked input characters and predicting the next sentence.

\subsubsection{Extend vocab size}

The pre-trained Chinese BERT\footnote[1]{https://github.com/google-research/bert} we used is trained by Google Inc using a large amount of unlabeled Chinese text data, and it is a base model, which has 12 encoder layers of transformer block, a hidden size of 768 and 12 attention heads. To predict the pronunciations of the SCPCs, we add the NCMCs into the vocabulary. The size of the pre-trained Chinese BERT vocabulary is 21,128. There are 354 polyphonic characters in our dataset, and we create 741 new monophonic characters and extend vocab size of the pre-trained Chinese BERT from 21,128 to 21,869.

\subsubsection{Initialization of new tokens}

Before we train the extended BERT model, the weights of 741 new tokens must be initialized. We can initialize a new token by copying any one of the old token's weights, but it usually turns out to be the SCPC or the [UNK]. After that, we get the Chinese polyphone BERT.

 \begin{table}[h]
	\centering
	\small
	\caption{\textit{Comparison between the pre-trained Chinese BERT and the Chinese polyphone BERT.}}
	\begin{tabular}{ccc}
		\toprule
		\textbf{Model}&\textbf{Vocab size}&\textbf{Params} \\
		\midrule
		Pre-trained BERT&21,128&102,290,312 \\
		Polyphone BERT&21,869&102,860,141 \\
		\bottomrule
	\end{tabular}
\end{table}

From Tabel 1, we know that the Chinese polyphone BERT is 0.55\% larger than the original BERT, which is acceptable. The extra 569,829 parameters appear in the embedding layer (741×768=569,088) and the MLM layers (741 bias).

\begin{figure}[h]
	\centering
	\includegraphics[width=\linewidth]{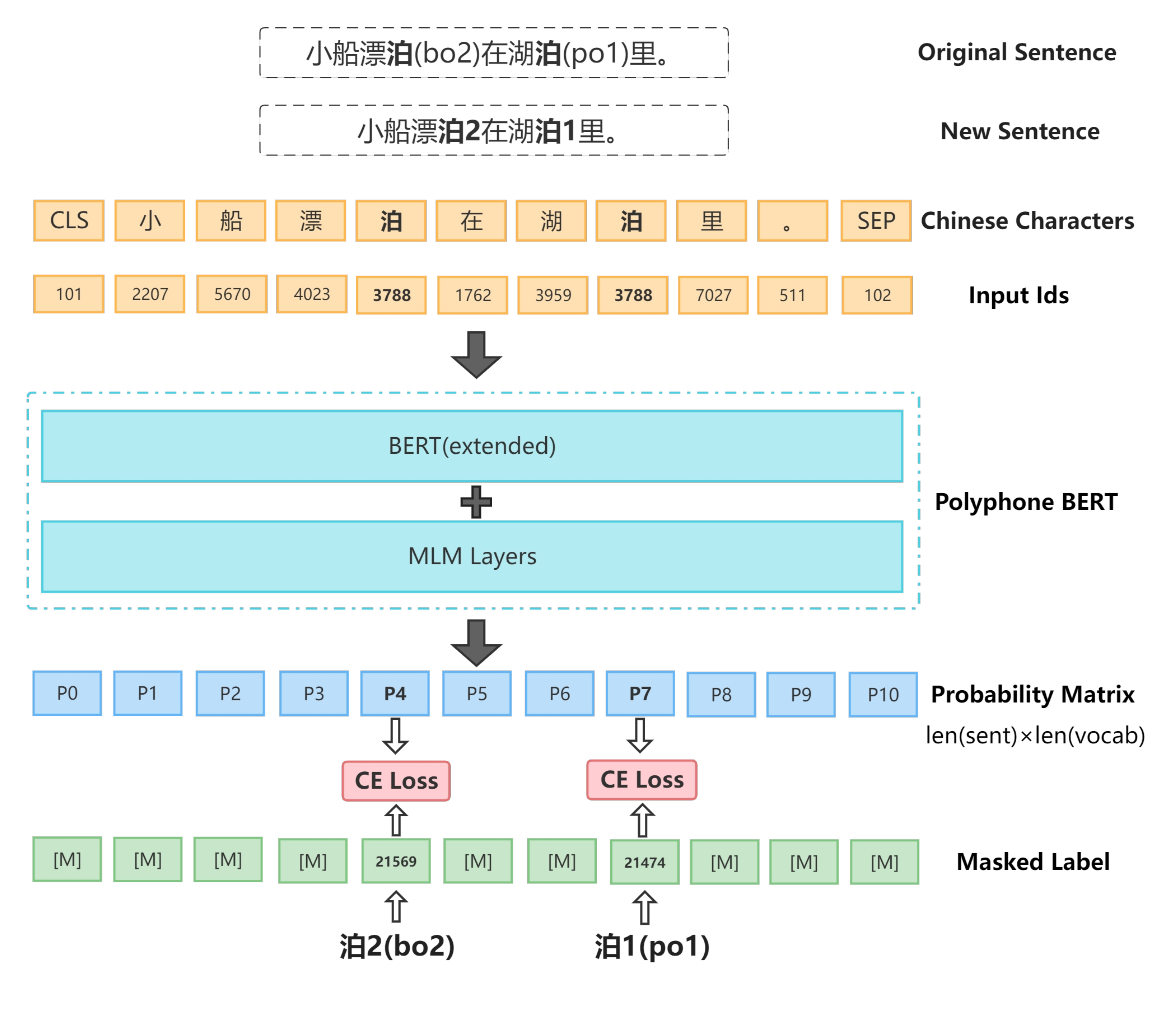}
	\caption{\textit{Training of a sample sentence.}}
\end{figure}

\subsection{Training details}

First of all, we turn all the training sentences, each one of which contains one SCPC at least, into new sentences with NCMCs. And then we train the Chinese polyphone BERT with the new sentences. The training of our method is a variant of predicting the masked input characters, the only differences are that we mask 100\% of the NCMCs and we use the tokens of SCPCs instead of the [MASK] tokens in the inputs.

As shown in Figure 4, an original sentence is turned into a new sentence by replacing the SCPC \begin{CJK}{UTF8}{gbsn}“泊”\end{CJK} with the NCMC \begin{CJK}{UTF8}{gbsn}“泊1”\end{CJK} and \begin{CJK}{UTF8}{gbsn}“泊2”\end{CJK}, and then we train the Chinese polyphone BERT with the new sentence. The NCMCs \begin{CJK}{UTF8}{gbsn}“泊2”\end{CJK} and \begin{CJK}{UTF8}{gbsn}“泊1”\end{CJK} are masked in the modified MLM task, and the [MASK] token is replaced by the SCPC \begin{CJK}{UTF8}{gbsn}“泊”\end{CJK}. The sparse categorical crossentropy loss is used in the training stage.

\subsection{Inference}

In the inference stage, the polyphone BERT accept a sentence with at least one SCPC as input and outputs a probability matrix, Figure 5 is an example of \begin{CJK}{UTF8}{gbsn} “小舟在湖中心漂泊”\end{CJK}(a boat drifted in the center of the lake). We can get the probability vector of the input SCPC \begin{CJK}{UTF8}{gbsn}“泊”\end{CJK} by index 8, and get the token ids of the NCMCs of the input SCPC \begin{CJK}{UTF8}{gbsn}“泊”\end{CJK}, which are 21,474 and 21,569. By comparing the probabilities of the token ids 21,474 and 21,569, we know the SCPC \begin{CJK}{UTF8}{gbsn}“泊”\end{CJK} of this sentence should be replaced by NCMC \begin{CJK}{UTF8}{gbsn}“泊2”\end{CJK} and its Pinyin should be \begin{CJK}{UTF8}{gbsn}“bo2”\end{CJK}.

 \begin{figure}[h]
  \centering
  \includegraphics[width=\linewidth]{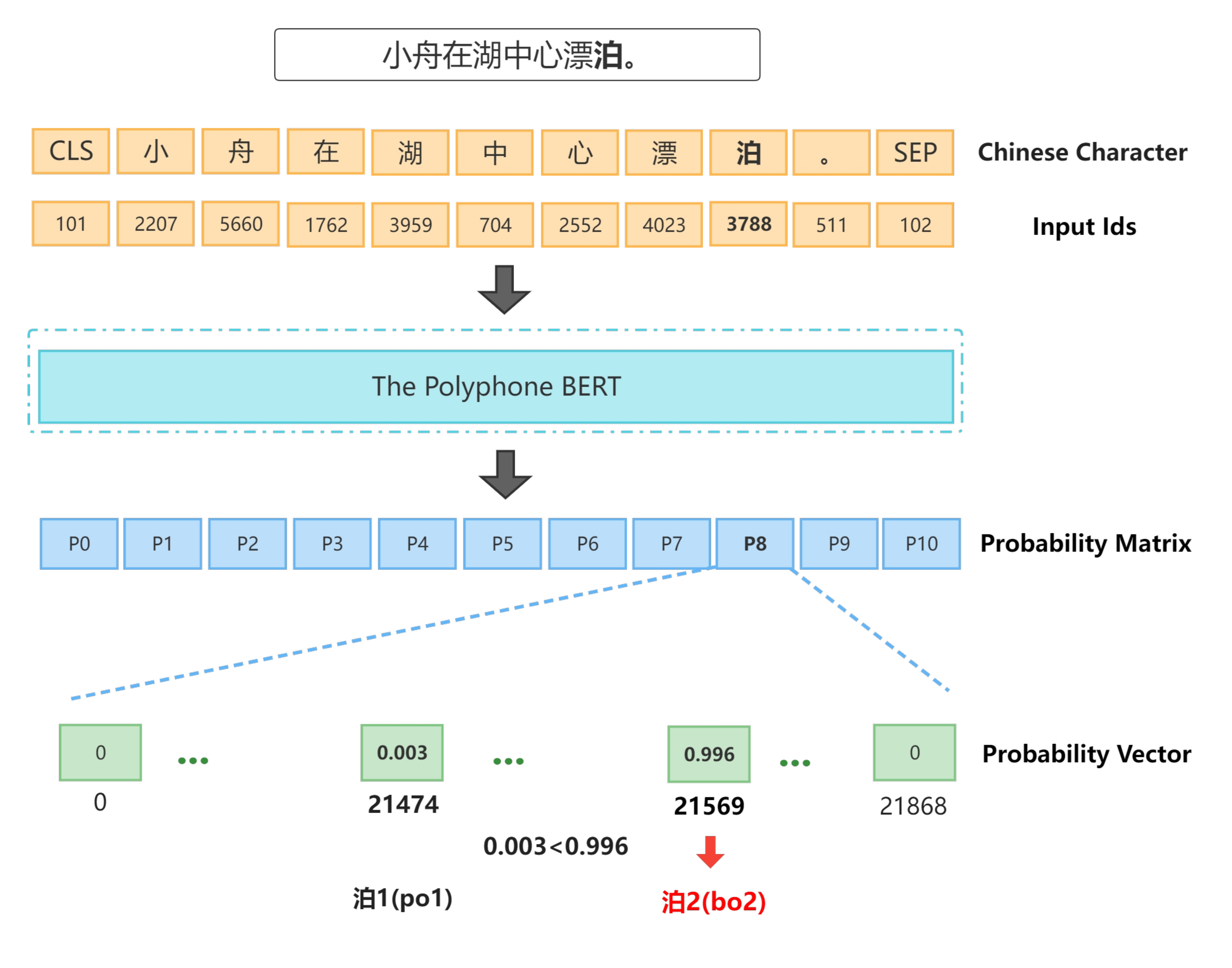}
  \caption{\textit{Inference of a sample sentence.}}
 \end{figure}

\section{Experiment}

\subsection{Dataset}

The experiments were conducted on one training set and four test sets. There are 1,419,803 sentences containing 354 most frequently used polyphonic characters in the training set. All the training data are selected from several web fiction and labeled manually. With the labeled Pinyin of the SCPCs in the training data, we turned the 354 SCPCs into 741 NCMCs.

 \begin{table}[h]
	\centering
	\small
	\caption{\textit{Statistics of the datasets used in our experiments.}}
	\begin{tabular}{ccc}
		\toprule
		&\textbf{sentences}&\ \\
		\midrule
		Training set&1,419,803 \\
		Test set 1&4,000\\
		Test set 2&2,475 \\
		Test set 3&5,502 \\
		Test set 4&100 \\
		\bottomrule
	\end{tabular}
\end{table}

To more fully compare the performance of different approaches, we collected 4 different test sets. Test set 1 contains 4,000 sentences coming from the same web fictions, and each Pinyin of the SCPCs occurs at least in 1,000 different sentences of the training set; Test set 2 contains 2,475 sentences in total, derived from another web fiction; Test set 3 contains 5,502 sentences selected from the test set of g2pM\footnote[2]{https://github.com/kakaobrain/g2pM}, in which many SCPCs appear as a part of a proper noun, making it difficult to predict the correct Pinyins for the SCPCs; Test set 4 contains 100 sentences, and in each one of them, a SCPC appears several times with different Pinyins, an example as \begin{CJK}{UTF8}{gbsn}“鱼拼命挣扎，鱼刺扎破了手，他随意包扎一下”\end{CJK}(The fish struggled desperately, and the fish pricked and broke his hand, he bandaged at will), and the correct Pinyins of the SCPC \begin{CJK}{UTF8}{gbsn}“扎”\end{CJK} in order is [zha2,zha1,za1]. Test set 4 is the most difficult of the four test sets.

\subsection{Experimental setting}
We implemented the following seven systems and used accuracy rate as evaluation criteria for comparison:
\begin{enumerate}
\item \textbf{BLSTM}: We use a 2 layer BLSTM with hidden size equal to 512, followed with two fully connected layers and a softmax layer to predict pronunciation. We split the corpus into mini batch with the batch size of 64.

\item \textbf{BERT(base, frozen)-FC}: Since we have a relatively large amount of training data, we set the parameters of BERT frozen and followed with two layers of fully-connected network, the hidden units of the first fully-connected layer is 768 with a dropout rate of 0.5 during training stage, unlike \cite{dai2019disambiguation}, we make the fully-connected network share parameters. During training stage, we adopted the Adam\cite{kingma2014adam} optimizer and set the learning rate to 5e-5. We split the corpus into mini batch with the batch size of 64.
\item \textbf{BERT(base, trainable)-FC}: Same as system 2 but set the parameters of the BERT model trainable.
\item \textbf{BERT(base, frozen)-BLSTM}: Same as system 2 but inserted 1 layer of BLSTM with hidden units equal to 512, we used the training strategy described in system 2.
\item \textbf{BERT(base, trainable)-BLSTM}: Same as system 4 but set the parameters of the BERT model trainable.
\item \textbf{Polyphone BERT(SCPC init)}: Our proposed model. NCMCs tokens are initialized by the weights of their SCPCs. We adopted the Adam optimizer and set the learning rate to 5e-6, and the batch size is 64.
\item \textbf{Polyphone BERT(UNK init)}: Same as system 6 but NCMCs tokens are initialized by the weights of UNK.
\end{enumerate}

\begin{table*}[th]
	\centering
	\small
	\caption{\textit{Accuracy on four test sets of different systems.}}
	\begin{tabular}{clccccc}
		\toprule
		\textbf{ID}&\textbf{System}&\textbf{Test set 1}&\textbf{Test set 2}&\textbf{Test set 3}&\textbf{Test set 4}&\textbf{Average} \\
		\midrule
		1&BLSTM&99.4&91.0&87.4&76.6&88.6 \\
		\midrule
		2&BERT(base, frozen)-FC&99.4&96.4&90.8&80.9&91.9 \\
		3&BERT(base, trainable)-FC&99.5&96.2&91.7&83.0&92.6 \\
		4&BERT(base, frozen)-BLSTM&99.5&96.4&90.5&80.9&91.8 \\
		5&BERT(base, trainable)-BLSTM&99.4&96.5&91.9&80.0&91.9 \\
		\midrule
		6&Polyphone BERT(SCPC init)&\textbf{99.9}&\textbf{96.9}&\textbf{93.0}&\textbf{87.2}&\textbf{94.3} \\
		7&Polyphone BERT(UNK init)&99.7&96.7&92.8&86.0&93.8 \\
		\bottomrule
	\end{tabular}
\end{table*}

\subsection{Results and analysis}

\subsubsection{Evaluation of different systems}

Table 3 presents the results of our experiments. We can observe that:

(1) Our proposed polyphone BERT achieves the best results on all four test sets. System 6 (polyphone BERT with SCPC init) shows an accuracy of nearly 100\% on test set 1, and at least get 1.1\% (from 91.9\% to 93.0\%) improvement of accuracy rate on test set 3, 4.2\% (from 83.0\% to 87.2\%) on test set 4 compared with other methods. The 87.2\% accuracy rate on test set 4 indicates that the polyphone BERT also has strong robustness. A possible reason is that polyphone BERT can get benefit from the pre-trained MLM layers of the original BERT. And turning the polyphone disambiguation into a pre-training process also makes a difference.

(2) The seven systems can be divided into three categories: the classifier model (system 1), the BERT-based classifier models (system 2 to system 5), and the polyphone BERT models (system 6 and system 7). The BERT-based classifier models get 3.5\% (from 88.6\% to 92.1\%) improvement of average accuracy compared with the classifier model, and the number turns out to be 2\% (from 92.1\% to 94.1\%) when comparing the polyphone BERT models with BERT-based classifier models. 

(3) All systems achieve an accuracy rate above 99\% on test set 1. Test set 1 was selected from the same web fiction with the training set. Furthermore, each Pinyin of the SCPCs in test set 1 occurs at least in 1,000 different sentences of the training set. This tells us that if we have enough training data, such as 1,000 different sentences of each Pinyin for the SCPCs, all systems could have a good performance.

(4) All systems get the worst performance on test set 4. It is a tough nut to crack when there is a polyphone character with different Pinyins in one sentence. The polyphone BERT models obtain 5.4\% (from 81.2\% to 86.6\%) respective accuracy improvement over the BERT-based classifier models on average. A possible reason is that BERT can't supply a distinguishable text representation for the classifiers in such a case.

\subsubsection{Impact of polyphone BERT initialization}

By comparing the results of system 6 and system 7, we can find that the initialization of new tokens will play an influence on the performance of the polyphone model, and system 6 gets better results on all four test sets. We initialize the polyphone BERT by these two methods and directly predict the Pinyin of the SCPC \begin{CJK}{UTF8}{gbsn}“泊”\end{CJK} in \begin{CJK}{UTF8}{gbsn} “小舟在湖中心漂泊”\end{CJK}(a boat drifted in the center of the lake) without training the models. The initial probabilities of the NCMCs \begin{CJK}{UTF8}{gbsn}“泊1”\end{CJK} and \begin{CJK}{UTF8}{gbsn}“泊2”\end{CJK} is shown in the Table 4. The result indicates that all NCMCs of the same SCPC have equal initial probabilities and the SCPC init method gets bigger initial probabilities for NCMCs, which means a smaller initial loss for training.

\begin{table}[h]
	\centering
	\small
	\caption{\textit{The initial probabilites.}}
	\begin{tabular}{ccc}
		\toprule
		&\textbf{\begin{CJK}{UTF8}{gbsn}泊1(po1)\end{CJK}}&\textbf{\begin{CJK}{UTF8}{gbsn}泊2(bo2)\end{CJK}} \\
		\midrule
		SCPC init&0.321&0.321 \\
		UNK init&8e-4&8e-4\\
		\bottomrule
	\end{tabular}
\end{table}

\subsubsection{Impact of freezing BERT parameters}

By comparing the results of system 2 and system 3, it was proved to be effective to finetune BERT in the polyphone disambiguation task for the BERT-FC model. By setting the parameters of the BERT trainable, the BERT-FC model can get better performance on three of the four test sets and get a 0.7\% improvement of average accuracy rate. When it comes to system 4 and system 5, finetuning the BERT does not make an obvious impact on the performance of the BERT-BLSTM model, they both have a higher rate of accuracy on two test sets and a nearly average accuracy rate.
Same as \cite{zhang2020unified}, BERT-FC is a better choice than BERT-BLSTM in the polyphone task when BERT parameters are trainable, and get the highest average accuracy rate besides the polyphone BERT.

\section{Conclusion}

In this paper, we propose a polyphone BERT for polyphone disambiguation in Mandarin Chinese. We get the polyphone BERT model by extending a pre-trained Chinese BERT with 741 new Chinese monophonic characters, created from 354 Chinese polyphonic characters. Polyphone disambiguation task is turned into a pre-training task of the Chinese polyphone BERT. We conduct experiments on a training set containing 1,419,803 sentences and four test sets, and our method obtain 2\% improvement of average accuracy compared with the BERT-based classifier model, which is the prior state-of-the-art in polyphone disambiguation. What's more, we believe the idea of our method can be applied to other disambiguation tasks in natural language processing.

\section{Acknowledgements}

We would like to thank Jianlin Su, the author of bert4keras\footnote[3]{https://github.com/bojone/bert4keras}. Bert4keras is an open-source library of Python, which makes it easier for us to perform our experiments, and its author also gave us some valuable advice for this work.

\bibliographystyle{IEEEtran}

\bibliography{main}


\end{document}